\documentclass[aps,prapplied,reprint,floatfix,citeautoscript,longbibliography,superscriptaddress]{revtex4-1}
\usepackage{graphicx,float}
\usepackage{amsmath,amssymb,mathrsfs,dcolumn}
\usepackage{gensymb}
\usepackage[usenames]{color}
\usepackage{subfigure}
\usepackage{ulem}
\usepackage{hyperref}
\hypersetup{colorlinks=true, linkcolor=blue, citecolor=red, urlcolor=blue}
\begin{document}
\title{Electrically switchable entanglement channel in van der Waals magnets}
\author{H. Y. Yuan}
\affiliation{Institute for Theoretical Physics, Utrecht University, 3584CC Utrecht, The Netherlands}
\author{Akashdeep Kamra}
\affiliation{Center for Quantum Spintronics, Department of Physics, Norwegian University of Science and Technology, NO-7491 Trondheim, Norway}
\author{Dion M. F. Hartmann}
\affiliation{Institute for Theoretical Physics, Utrecht University, 3584CC Utrecht, The Netherlands}
\author{Rembert A. Duine}
\affiliation{Institute for Theoretical Physics, Utrecht University, 3584CC Utrecht, The Netherlands}
\affiliation{Center for Quantum Spintronics, Department of Physics, Norwegian University of Science and Technology, NO-7491 Trondheim, Norway}
\date{\today}

\begin{abstract}
Two dimensional layered van der Waals (vdW) magnets have demonstrated their potential to study both fundamental and applied physics due to their remarkable electronic properties.  However, the connection of vdW magnets to spintronics as well as quantum information science is not clear. In particular, it remains elusive whether there are novel magnetic phenomena only belonging to vdW magnets, but absent in the widely studied crystalline magnets. Here we consider the quantum correlations of magnons in a layered vdW magnet and identify an entanglement channel of magnons across the magnetic layers, which can be effectively tuned and even deterministically switched on and off by both magnetic and electric means. This is a unique feature of vdW magnets in which the underlying physics is well understood in terms of the competing roles of exchange and anisotropy fields that contribute to the magnon excitation. Furthermore, we show that such a tunable entanglement channel can mediate the electrically controllable entanglement of two distant qubits, which also provides a protocol to indirectly measure the entanglement of magnons. Our findings provide a novel avenue to electrically manipulate the qubits and further open up new opportunities to utilize vdW magnets for quantum information science.
\end{abstract}

\maketitle
{\it Introduction.---} Two dimensional (2D) van der Waals (vdW) magnets have attracted significant research interests for their remarkable features that include rich physical phenomena in the 2D limit and desirable tunability of magnetic properties by gate voltage \cite{Gong2017,Bevin2017,Burch2018,Gib2019}. Of particular interest is the interplay of vdW magnets with traditional spintronics which manipulates spin transport controlled by a wide range of knobs including magnetization dynamics, magnetic field, microwave, charge current, thermal gradient, and mechanic strain. Even though spin-orbit torque induced magnetization switching \cite{WangX2020,Dolui2020} and thermal driven magnon transport in 2D magnets was recently demonstrated \cite{Xing2019,Liu2020,Gupta2020,Wang2020}, resembling the observation in bulk magnets, it remains elusive whether there are novel magnetic phenomena only belonging to the vdW magnets and whether they are advantageous to advance the current horizon of spintronics. Here we will address this issue by considering the quantum correlations of magnons in a vdW magnet and its potential application in quantum information science. Primary opportunities of 2D materials have emerged in designing quantum-dot qubits, superconducting qubits, and topological quantum computing platforms \cite{Liu2019}. As we focus on 2D vdW magnets, however, the situation becomes different. Here the elementary excitation in the magnets are magnons with a continuous spectrum instead of qubits with discrete energy levels, and thus it becomes more natural to integrate with continuous variable quantum information \cite{Braun2005}, similar to the role of mechanical modes in optomechanics \cite{Markus2014}. Generating and manipulating quantum entanglement of continuous quantum variables is an essential step in this direction, and it also motivates our current work.

In this paper, we study the entanglement of magnons in a layered vdW magnet and identify an entanglement channel, robust against the magnetic dissipation. The magnon entanglement through this channel can be effectively tuned by both magnetic and electric fields. It is a unique feature of vdW magnets with comparable strength of exchange and anisotropy fields, and is absent in most bulk magnets. We further show that such an entanglement channel can bridge the entanglement of two qubits and it can be electrically switchable by tuning the gate voltage. Our results may significantly extend the research horizon of vdW magnets and promote the application of vdW magnets in quantum information science.

\begin{figure}
  \centering
  \includegraphics[width=0.45\textwidth]{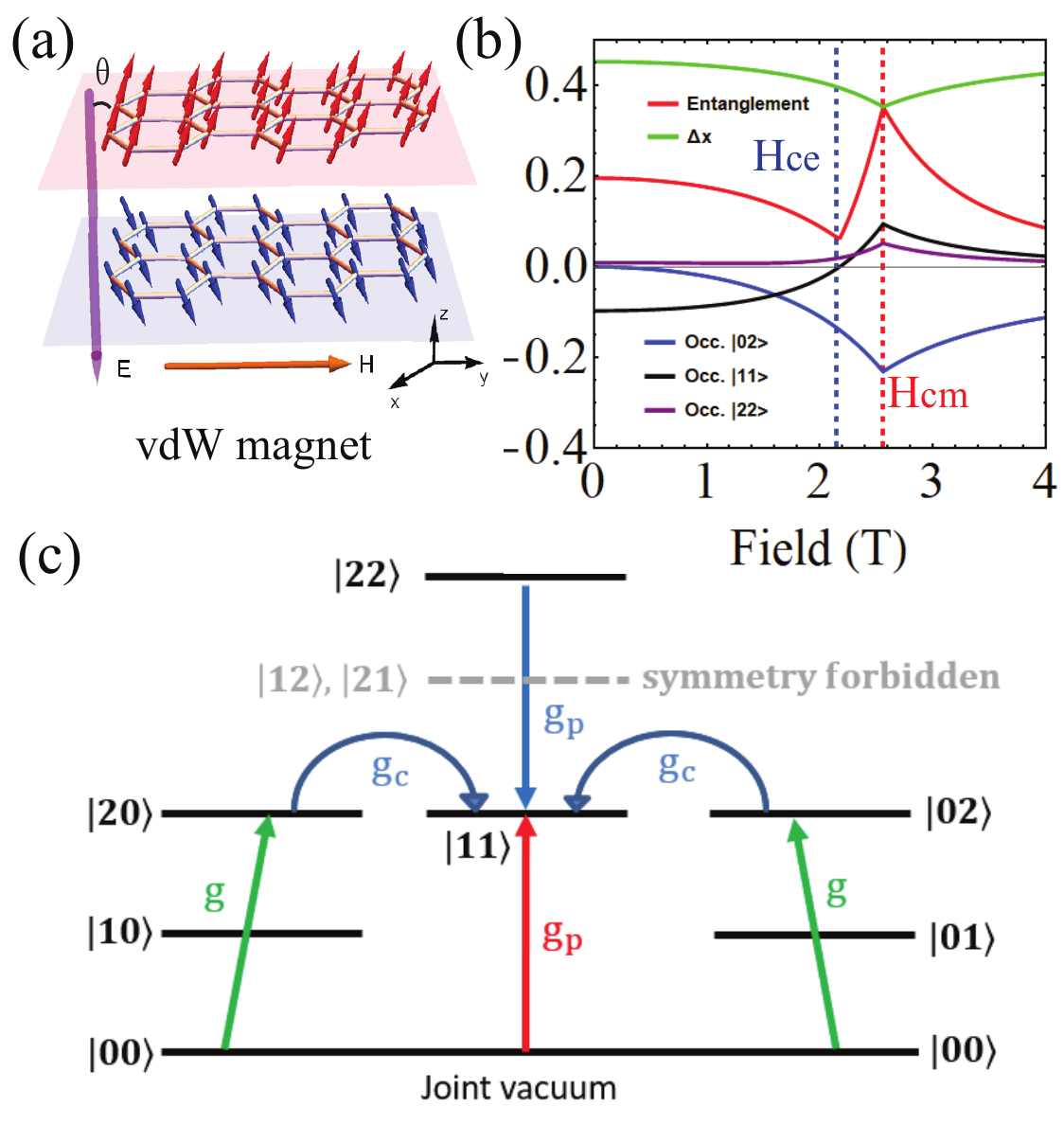}\\
  \caption{(a) Schematic illustration of the bilayer van der Waals magnets subject to an in-plane magnetic field ($\mathbf{H}$) and a normal electric field ($\mathbf{E}$) (b) Entanglement of magnons and occupation of Fock space in the ground state as a function of magnetic field. The magnetic parameters of $\mathrm{CrI_3}$ \cite{Zhang2020} are used with $2J=0.83$ T, $2K = 1.73$ T. The red dashed line indicates the magnetic transition field at $H_{\mathrm{cm}} = 2.46$ T and blue dashed line the local minimum of entanglement at $H_{\mathrm{ce}} = 2.16$ T. (c) Energy levels of the two magnon system. The occupation of $|11\rangle$ is crucial to determine the two magnon entanglement.}\label{fig1}
\end{figure}

{\it Physical model.---}Let us consider a layered vdW magnet shown in Fig. \ref{fig1}(a), described by the following Hamiltonian,
\begin{equation}
\mathcal{H}=\sum_{ij,\mathbf{R},\mathbf{R}'} J_{ij} \mathbf{S}_{i,\mathbf{R}}^{} \cdot \mathbf{S}_{j\mathbf{R}'} - K \sum_{i,\mathbf{R}} (S_{i,\mathbf{R}}^z)^2 - \mathbf{H} \cdot \sum_{i,\mathbf{R}} \mathbf{S}_{i,\mathbf{R}},
\end{equation}
where $i,j\in \{1,2\}$ are layer indices, $\mathbf{R}$ is the position of magnetic atoms, while the diagonal and off diagonal elements of exchange matrix $J_{ij}$ represent the intralayer and interlayer nearest neighbor exchange coupling, respectively. Furthermore, $K>0$ is the anisotropy coefficient and $\mathbf{H}$ is the applied magnetic field. The strength of both exchange and anisotropy fields is tunable by modifying the density of states of electrons through electric fields. Since the intralayer exchange coupling is much larger than the interlayer coupling ($J_{12} \ll J_{11},J_{22}$), the magnetization in each layer usually behaves like a macrospin. Following the experimental setup realized for $\mathrm{CrI_3}$ \cite{Zhang2020}, we consider an in-plane field $\mathbf{H}=He_y$.
Then the classical ground state of the system is a canted state $\theta =\arcsin H/H_{\mathrm{cm}}$ when $H < H_{\mathrm{cm}}$ and a collinear state $\theta =\pi/2$ when $H \ge H_{\mathrm{cm}}$, where the magnetic transition field $H_{\mathrm{cm}}\equiv2(J+K)S$ with $J_{12}=J_{21}\equiv J/2$ \cite{yuan2021power}. To consider the magnon excitation on this ground state, we perform a Holstein-Primakoff transformation \cite{HP1940} near the classical ground state and derive the effective Hamiltonian of the uniform precession mode as,
\begin{equation}
\begin{aligned}
\mathcal{H}&=\sum_{i=1}^2 \omega_0 a_i^\dagger a_i + \frac{1}{2}g(a_i^\dagger a_i^\dagger + a_i a_i) \\
&+g_c(a_1^\dagger a_2 + a_1 a_2^\dagger)+g_p(a_1^\dagger a_2^\dagger + a_1 a_2),
\end{aligned}
\label{ham}
\end{equation}
where $\omega_0=(J + K)S \cos 2\theta + KS \cos^2\theta + H \sin \theta, g = KS\sin^2\theta, g_p = JS \cos^2\theta,
g_c=JS\sin^2\theta$, and $a_i$ ($a_i^\dagger$) is the magnon annihilation(creation) operators on $i-$th magnetic layer.

{\it Ground state.---} Let us first study the ground state properties, which will be useful to understand the quantum correlations of magnons subject to dissipation. By numerically solving the Schr\"{o}dinger equation $\mathcal{H} | \psi \rangle = E | \psi \rangle$ in the bipartie Fock basis, where $| \psi \rangle = \sum_{m,n=1}^N C_{mn} |mn\rangle $, and $|mn\rangle$ refers to the occupation of the Fock state with particle number $m$ and $n$ in layer 1 and 2, respectively, we can derive energy and wavefunction of the ground state after taking a proper truncation of the maximum number of magnon excitations ($N$). In general, one may quantify the strength of entanglement using the von Neumann entropy, Duan-Simon criteria \cite{Duan2000,Simon2000} and log-negativity \cite{Adesso2007}. To be consistent with the discussion on the mixed state case presented below, we choose log-negativity, where a state with zero log-negativity is separable and that with larger log-negativity has stronger entanglement. A well-known example is that the log-negativity of a two-mode squeezed state is equal to its squeezing parameter. The more squeezing of the state in position or momentum dimension, the stronger the entanglement of the two modes is.
To calculate the log-negativity in the magnonic systems, we first derive the covariance matrix (CM) of the system in the form
$\mathbf{V}_{ll'}=(\langle u_l u_{l'} \rangle + \langle u_{l'} u_l\rangle$)/2, where $\mathbf{u}=(x_1,p_1,x_2,p_2)$, $\langle u_l u_{l'} \rangle= \langle \psi | u_i u_j| \psi \rangle$ and the quadrature operators are defined as $x_i=(a_i + a_i^\dagger)/\sqrt{2}$, $p_i=-i(a_i - a_i^\dagger)/\sqrt{2}$.
Due to the permutation symmetry of the system ($1\leftrightarrow 2$), the CM should take the form $\mathbf{V}_{4\times 4}=\left(
  \begin{array}{cc}
    \mathbf{A}_{2\times 2} & \mathbf{C}_{2\times 2} \\
    \mathbf{C}^T_{2\times 2} & \mathbf{A}_{2\times 2} \\
  \end{array}
\right)$,
then the logarithmic negativity is calculated as $E_N=\max[0,-\ln[2\eta^-]]$, where $\eta^-=\sqrt{\sum \mathbf{V}-\sqrt{\sum \mathbf{V}^2-4\det \mathbf{V}}}$ with $\sum \mathbf{V}=2\det(\mathbf{A})-2\det(\mathbf{C})$ and $\det (\mathbf{V})$ being the two sympletic invariants of $\mathbf{V}$ \cite{Adesso2007}.

Figure \ref{fig1}(b) shows that, with the increase of magnetic field, the entanglement first drops until a field $H_{\mathrm{ce}}$ before the magnetic transition and then rises to a peak at the transition ($H_{\mathrm{cm}}$) (red dashed line). As a comparison, the two mode squeezing $\Delta x = \sqrt{\langle (x_1+x_2)^2\rangle }$ takes on a single minimum at $H_{\mathrm{cm}}$ (green line). In the case of the wavefunction being described by a two-mode squeezed state, the entanglement becomes a monotonous function of the squeezing parameter \cite{Adesso2007,Yuan2020afm,Akash2019}. In contrast, our system deviates from such a wavefunction and thus admits additional entanglement channels as discussed further below.


To understand the essential physics, we first analyze the symmetry of the Hamiltonian. Both the on-site squeezing $a_i^\dagger a_i^\dagger$  and inter-site squeezing $a_1^\dagger a_2^\dagger$ only allow an even number of excitations ($|00\rangle, |02\rangle, |11\rangle, |20\rangle,...$) such that the ground state wave function can be written as $| \psi \rangle = C_0|00\rangle +C_2|02\rangle C_2 + C_3 |11\rangle + C_2|20\rangle +C_4 |22\rangle$, where we have truncated the Hilbert space at $N=2$, as we shall see, which is sufficient to capture the qualitative physics. The entanglement of such a state means that it cannot be factorized as $| \psi \rangle =| \psi_1 \rangle \otimes |\psi_2\rangle $, where $| \psi_i \rangle = c_0 |0\rangle+ c_1|1\rangle + c_2 |2\rangle$. This factorization is straightforward when $c_1=0$. Given $c_1 \neq 0$, this factorization is impossible for the inevitable appearance of odd excitation ($|10\rangle , |01\rangle...$). Hence it is expected that the occupation of $|11\rangle$ ($c_1$) is crucial to determine the entanglement of the ground state. On the other hand, as shown in Fig. \ref{fig1}(c), the occupation of $|11\rangle$ acquires contributions from two competing effects: (i) The direct parametric excitation ($g_p$), whose strength decreases monotonically as $\theta$ increase from zero to $\pi/2$. (ii) The coherent transfer of $|11\rangle$ to $|02\rangle$ and $|20\rangle$ ($g_c$) followed by an on-site parametric process to  $|00\rangle$ ($g$) and the reversal process, whose strength increases as $\theta$ increases. These two channels interfere destructively and result in a minimal entanglement state at a critical value $H_{\mathrm{ce}}$ before $H_{\mathrm{cm}}$. Above $H_{\mathrm{ce}}$, the coherent channel dominates the entanglement and it keeps increasing up to the magnetic transition field $H_{\mathrm{cm}}$. As we increase the fields further, $g_c$ does not change any longer, while the gap between $|00\rangle$ and $|02\rangle$ keeps increasing to suppress the on-site parametric excitation. Hence the occupation of $|11\rangle$ and the ground state entanglement decrease. The numerical calculation of the particle occupation in Fock space as a function of field in Fig. \ref{fig1}(b) shows that the minimum entanglement indeed happens when the occupation of $|11\rangle$ is minimum, which is consistent with the intuitive picture.

{\it Role of dissipation.---} Now we generalize this picture to a system with dissipation, which will mix the contributions from ground and excited states and stabilize the system in a steady state. In general, the steady state described by a density matrix can be characterized by its first and second moments, i..e $\langle a_i \rangle$ and $\langle a_i a_j \rangle$. For the lack of coherent pumping in Eq. (\ref{ham}),  $\langle a_i \rangle=0$, while $\langle a_i a_j \rangle$ can be calculated by either solving the Lindblad equation or Lyapunov equation $\mathbf{MV}+\mathbf{VM}=-\mathbf{D}$ \cite{Vitali2007}. Here $\mathbf{D}=diag(\gamma_1,\gamma_1,\gamma_2,\gamma_2)$ and the drift matrix of the system ($d\mathbf{u}/dt=\mathbf{M} \cdot \mathbf{u}$) reads,
\begin{equation}
\mathbf{M}=\left (
\begin{array}{cccc}
  -\gamma_1 & -g+\omega_0 & 0 & g_c-g_p \\
  -g-\omega_0 & -\gamma_1 & -g_c-g_p& 0 \\
  0 & g_c-g_p & -\gamma_2 & -g+\omega_0 \\
  -g_c-g_p & 0 & -g - \omega_0 & -\gamma_2
\end{array}
\right).
\end{equation}
where $\gamma_1$ and $\gamma_2$ are the dissipation coefficients of the two magnetic layers.

\begin{figure}
  \centering
  \includegraphics[width=0.48\textwidth]{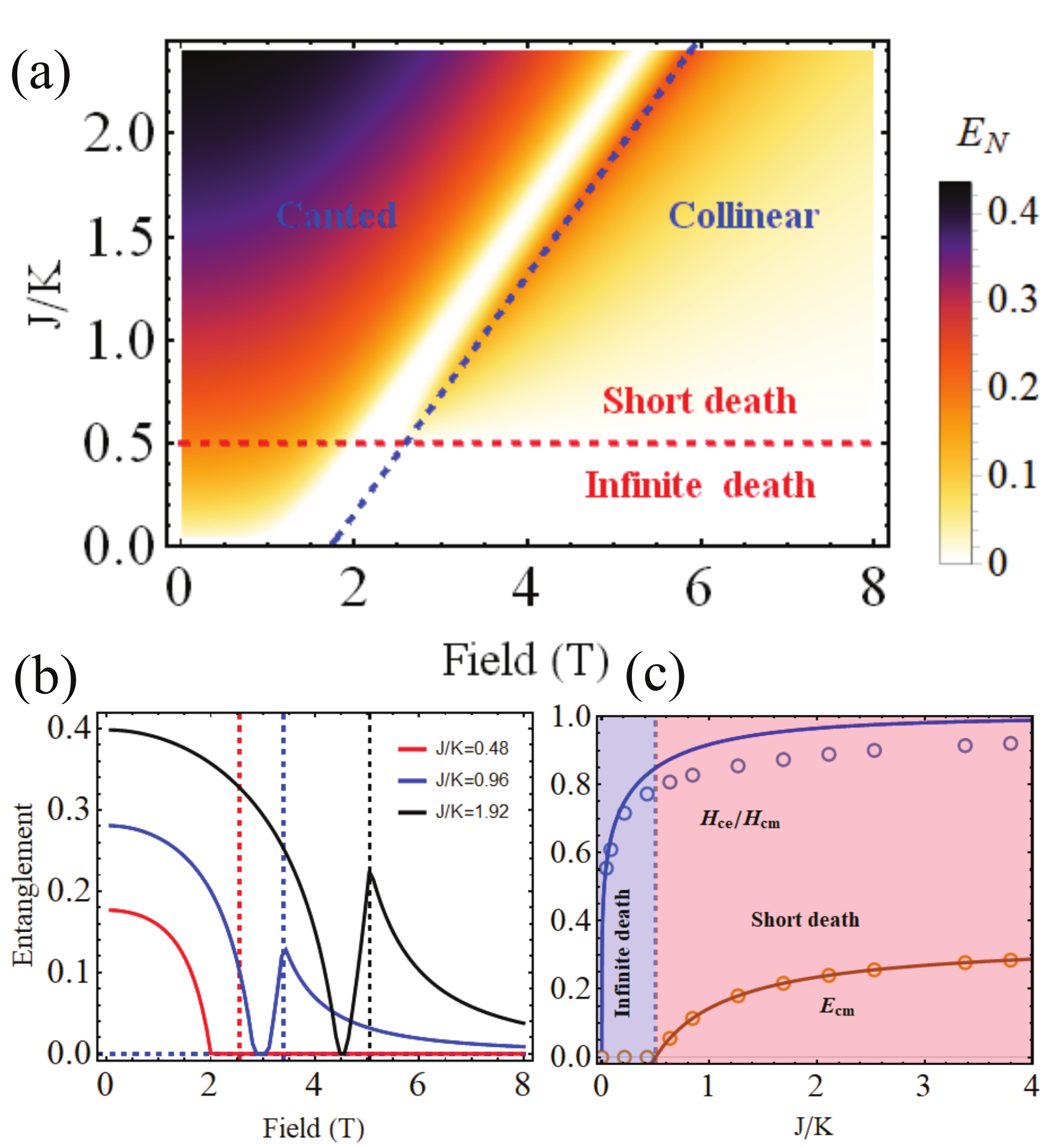}\\
  \caption{(a) Steady magnon-magnon entanglement in the $(H,J/K)$ plane. (b) Steady entanglement as a function of external field for $J/K=$0.48 (red line), 0.96 (blue line) and 1.92 (black line), respectively. The dashed lines indicate the transition fields $H_{\mathrm{cm}}$ locating at $2(J+K)S$. $\gamma_1=\gamma_2=0.01$. (c) Entanglement at the transition field ($E_{\mathrm{cm}}$) and critical field to have zero entanglement ($H_{\mathrm{ce}}/H_{\mathrm{cm}}$) as a function of $J/K$. The symbols are full numerical results based on Hamiltonian (\ref{ham}) and lines depict Eqs. (\ref{hce}) and (\ref{Ecm}).}\label{fig2}
\end{figure}

Then we can again quantify the entanglement by calculating the log-negativity.
Figure \ref{fig2}(a) shows the magnon-magnon entanglement in the plane of ($H$, $J/K$). Depending on the values of $J/K$, two regimes can be classified, as shown in Fig. \ref{fig2}(b) which is extracted from the phase diagram for $J/K=0.48$ (red line), 0.96 (blue line) and 1.92 (black line), respectively. For $J/K<0.5$, the entanglement monotonically decreases to zero at a critical field $H_{\mathrm{ce}}<H_{\mathrm{cm}}$ and never revives again. While for $J/K>0.5$, the entanglement first dies at $H_{\mathrm{ce}}$, revives after a finite field range, and further reaches a local maximum at the transition field $H_{\mathrm{cm}}$. Such a field dependence resembles the ground state case, as discussed in Fig. \ref{fig1}.
This similarity suggests the coherent channel and parametric channel still manifest themselves in the steady state. However, the coherent channel is more easily to be switched off under the influence of dissipation at smaller $J$. To illustrate this point, we discuss these regimes in detail as follows.

 Let us first study the low field regime by setting $g_c=0$. Then we can analytically derive the CM and calculate the magnon-magnon entanglement as,
\begin{equation}
E_N=\max\left (0,\frac{1}{2} \ln \frac{\left [(\omega_0 + g_p)^2-g^2\right ]\left [(\omega_0 - g_p)^2-g^2\right ]}{\left(\omega_0^2-g_p\sqrt{\omega_0^2+\gamma^2}\right )^2-g (\omega_0^2+\gamma^2)} \right ).
\end{equation}
Since the dissipation is usually sufficiently smaller than the on-site frequency ($\gamma \ll \omega_0$)\cite{notedissipation}, $E_N$ can be further simplified as,
\begin{equation}
E_N=\max(0,\ln \sqrt{(\omega_0+g_p)^2 - g^2} - \ln \omega_0).
\end{equation}
One immediately sees that the entanglement positively depends on the parametric excitation of two types of magnons ($g_p$), and there exists a critical condition at witch the steady entanglement is zero, i.e.,
\begin{equation}
g_p^2+2\omega_0g_p-g^2=0.
\end{equation}
After substituting the field dependence of $g,g_p,\omega_0$, we can calculate the critical magnetic field,
\begin{equation}
H_{\mathrm{ce}}=H_{\mathrm{cm}} \sqrt{\frac{3J/K + 4}{2J/K +3 + \sqrt{J^2/K^2 + 2J/K + 4 +4K/J}}}.
\label{hce}
\end{equation}
It is straightforward to verify that $H_{\mathrm{ce}}$ is a monotonic increasing function of $J/K$ and it approaches 0, when $J/K \rightarrow 0$, and $H_{\mathrm{cm}}$ when $J/K \rightarrow \infty$. This prediction describes the numerical results quite well for small $J/K$, as shown in Fig. \ref{fig2}(c). As $J/K$ increases, the critical field approaches $H_{\mathrm{cm}}$, and the influence of the coherent process needs to be included.

In the coherent regime ($g_p=0$), the CM of the system is also analytically solvable, and the log-negativity reads,
\begin{equation}
E_N=\max \left(0,\frac{1}{2}\ln \frac{\omega_0^2-(g-g_c)^2}{\omega_0^2-g_c^2} \right ).
\end{equation}

This result implies that the magnons are not entangled for $g> 2g_c$ (or equivalently  $K>2J$), which can be easily satisfied in vdW magnets, but not favorable in crystalline magnets ($J \gg K$). This readily explains the infinite death of entanglement without revival observed in Fig. \ref{fig2}(a). Here the tunability of $J/K$ through electric field enables an electrically switchable entanglement channel as we shall see below. In the regime $g<2g_c$ ($K<2J$), there will be a finite entanglement between magnons near the magnetic transition field. Depending on the magnitude of applied field, two regimes can be classified. (i) When $H\geq H_{\mathrm{cm}}$, $\theta=\pi/2$,
\begin{equation}
E_N^>=-\frac{1}{2} \ln \left (1 - \frac{(2J-K)KS^2}{(H-2KS)(H-2JS)}\right ),
\label{Ecm}
\end{equation}
which decreases monotonically as field increases and the maximum entanglement at $H=H_{\mathrm{cm}}$ is $E_{\mathrm{cm}}=\ln(\sqrt{1+K/2J}/\sqrt{2})$.
(ii) When $H<H_{\mathrm{cm}}$, $\theta = \arcsin H/H_{\mathrm{cm}}$, the log-negativity can be rewritten as,
\begin{equation}
E_N^<=\sqrt{\frac{a'_4 H^4 + a_2 H^2 + a_0}{a_4 H^4 + a_2 H^2 + a_0}},
\end{equation}
where $a'_4= (J+K)(J-K),a_4=J(J-2K), a_2=8KS^2(J+2K)(J+K)^2, a_0=-16S^4(J+2K)^2(J+K)^4$. Note that $a'_4>a_4$ due to $K<2J$, this implies that the entanglement will increase with the external field and reach its maximum value $E_{\mathrm{cm}}$ at $H_\mathrm{cm}$. The prediction of this maximum entanglement captures the numerical result perfectly as shown in Fig. \ref{fig2}(c) (brown line).

\begin{figure}
  \centering
  \includegraphics[width=0.45\textwidth]{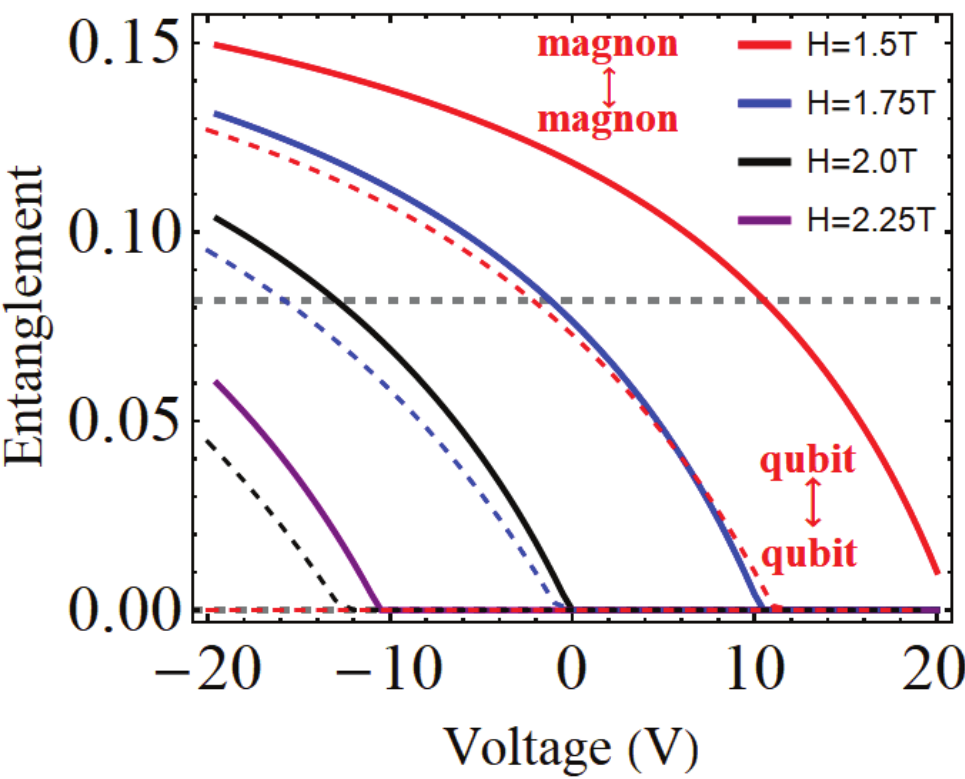}\\
  \caption{Entanglement of magnons (solid lines) and qubits (dashed lines) as a function of gate voltage for $H=1.5$ T (red line), 1.75 T (blue line), 2.0 T (black line) and 2.25 T (purple line), respectively. The gray line at $E_N=0.085$ is a threshold of magnon-magnon entanglement to switch on and off the qubit-qubit entanglement. Parameters are $J_0=0.865 ~\mathrm{T}, b_J=9.5\times 10^{-3} ~\mathrm{T/V}, K_0=0.415 ~\mathrm{T}, b_K=5.5 \times 10^{-3} ~\mathrm{T/V}$ \cite{Zhang2020}. $\omega_1=-\omega_2=0.01 ~\mathrm{T}, g_{\mathrm{qm}}=10^{-4} ~\mathrm{T}$ \cite{Bertelli2020, notegm}.}\label{fig3}
\end{figure}

{\it Qubit-qubit entanglement.---} We have demonstrated a tunable and robust entanglement channel in bilayer vdW magnets by magnetic field. Furthermore, this quantum channel can also be turned on and off by electric field through the modification of exchange and anisotropy field. With bilayer $\mathrm{CrI_3}$ as an example, recent experiments show that both the exchange and anisotropy fields decrease with the gate voltage $V$ as $J=J_0-b_JV$ and $K=K_0-b_KV$ \cite{Zhang2020}. By taking this dependence into account, we  numerically calculate the two magnon entanglement as a function of gate voltage at different values of magnetic fields in Fig.  \ref{fig3} (solid lines). Clearly, the gate voltage can be an effective knob to switch the entanglement of magnons. This is a unique feature of vdW magnets, because the magnitude of anisotropy is usually three orders smaller than the exchange field in typical crystalline antiferromagnets and its change by electric field will not have significant influence on the entanglement channel.

This tunable entanglement provides an avenue to manipulate two qubits electrically.  As a proof of concept, we consider two qubits coupled with the two layers of the vdW magnets respectively through dipolar interactions. Under the rotating wave approximation, the total Hamiltonian is written as \cite{notek},
\begin{equation}
\mathcal{H}_t= \mathcal{H} + \sum_{i=1}^2 \omega_i \sigma_i^z+ g_{qm}\sum_{i=1}^2(\sigma_i^+ a_i + \sigma_i^- a_i^\dagger),
\end{equation}
where $\sigma_i^z, \sigma_i^\pm$ are the Pauli matrices describing the $i-$th qubit, $\omega_i$ is the resonance frequency of the qubit, $g_{\mathrm{qm}}$ is the coupling strength. Here we first solve the steady density matrix of the whole system $\rho$ numerically and then calculate the reduced density matrix of the two qubits by tracing out the magnon degree of freedom, i.e., $\rho_{12}=tr_{a_1a_2}(\rho)$. Then we can quantify the qubit-qubit entanglement by concurrence defined as $C_{12}=\max (0,\lambda_1-\lambda_2-\lambda_3-\lambda_4)$,
where $\lambda_{1,2,3,4}$ are the square roots
of the eigenvalues of $\rho_{12}[ (\sigma^y \otimes \sigma^y)
\rho^*_{12} (\sigma^y \otimes \sigma^y)]$ in a non-increasing order \cite{Wooters1998,Yuan2018,Zou2020}. Figure \ref{fig3} shows that the concurrence is also tunable by gate voltage (dashed lines) and follows a similar trend as magnon entanglement (solid lines). However, there exists a threshold entanglement of magnons near 0.082, only above which the magnonic channel can mediate the entanglement of two qubits. To realize this proposal, one may optimize the material properties of vdW magnets and identify those with smaller exchange and anisotropy such that it could readily be integrated with the mature qubit platforms including nitrogen vacancy centers and superconducting qubits \cite{Wendin2017,Casola2018}. As an alternate, one may find a quantum platform working at the strong field regime, which has already been indicated in the recent experiments \cite{Kroll2018,Luthi2018}.

{\it Discussion and conclusion.---} In conclusion, we have shown that magnons in a layered vdW magnet are strongly entangled, and that this entanglement is subject to a sudden death before the magnetic transition field. The essential physics is ascertained to be the competing influence of parametric interaction of the two types of magnons simultaneously and the coherent transfer between the magnon states. On the practical side, our observation enables both magnetic and electrical means to switch on and off the entanglement channel and further makes it possible to utilize such a channel to bridge the entanglement of two or more qubits. These features are absent in the normal crystalline magnet because of the smallness of anisotropy compared with exchange field and limited tunability range of the magnetic parameters by electric fields.

\begin{acknowledgments}
 H.Y.Y acknowledges the European Union's Horizon 2020 research and innovation programme under Marie Sk{\l}odowska-Curie Grant Agreement SPINCAT No. 101018193. R.A.D. has received funding from the European Research Council (ERC) under the European Union's Horizon 2020 research and innovation programme (Grant No. 725509). A.K. was funded by the Research Council of Norway through its Centers of Excellence funding scheme (Project No. 262633 ``QuSpin").
\end{acknowledgments}


\begin{thebibliography}{99}
\bibitem{notedissipation} As the dissipation increases to be comparable with the self-frequency ($\gamma \sim \omega_0$), the magnitude of
entanglement will decrease, while the critical field $H_{\mathrm{ce}}$ still exists. This regime is out of consideration in the current work.
\bibitem{notek} In general, the spin waves of all the wavevectors may mediate the coupling of two qubits. Here we consider the low temperature case and thus expect the dominate contribution comes from the uniform-precession mode with zero wavevector.
\end{thebibliography}
\end{document}